\documentclass[aps,pof,floatfix,twocolumn,superscriptaddress]{revtex4} 
\usepackage[latin1]{inputenc}
\usepackage[english]{babel}
\usepackage[T1]{fontenc}

\usepackage{textcomp}
\usepackage{graphicx}
\usepackage[]{amsmath}
\usepackage{amssymb}
\usepackage{color}

\begin{document}

\title{Chaotic mixing in effective compressible flows.}


\author{R. Volk}
\affiliation{Laboratoire de Physique de l'ENS de Lyon, CNRS UMR5672 et Universit\'e de Lyon, France}
\author{C. Mauger}
\affiliation{LMFA, CNRS UMR5509, \'Ecole Centrale Lyon, INSA de Lyon et Universit\'e de Lyon, France}
\author{M. Bourgoin}
\affiliation{LEGI, CNRS UMR5519, Universit\'e Joseph Fourier, Grenoble INP, France}
\author{C. Cottin-Bizonne}
\affiliation{Institut Lumi\`ere Mati\`ere, CNRS UMR5306, Universit\'e Claude Bernard Lyon 1, Universit\'e de Lyon, France}
\author{C. Ybert}
\affiliation{Institut Lumi\`ere Mati\`ere, CNRS UMR5306, Universit\'e Claude Bernard Lyon 1, Universit\'e de Lyon, France}
\author{F. Raynal}
\affiliation{LMFA, CNRS UMR5509, \'Ecole Centrale Lyon, INSA de Lyon et Universit\'e de Lyon, France}

\email{romain.volk@ens-lyon.fr}

\begin{abstract}
{We study numerically joint mixing of salt and colloids by a chaotic velocity field $\mathbf{V}$, and how salt inhomogeneities accelerate or delay colloid mixing by inducing a velocity drift $\mathbf{V}_{\rm dp}$ between colloids and fluid particles as proposed in recent experiments \cite{Deseigne2013}.
{We demonstrate that because the drift velocity is no longer divergence free, small variations to the total velocity field drastically affect the evolution of colloid variance $\sigma^2=\langle C^2 \rangle - \langle C \rangle^2$. A consequence is that mixing strongly depends on the mutual coherence between colloid and salt concentration fields, the short time evolution of scalar variance being governed by a new variance production term $P=- \langle C^2 \nabla \cdot \mathbf{V}_{\rm dp} \rangle/2$ when scalar gradients are not developed yet so that dissipation is weak.}
Depending on initial conditions, mixing is then delayed or enhanced, and it is possible to find examples for which the two regimes (fast mixing followed by slow mixing) are observed consecutively when the variance source term reverses its sign. This is indeed the case for localized patches modeled as gaussian concentration profiles.}
\end{abstract}
\date{\today}
\maketitle

{Mixing of a scalar field $C$ by chaotic flows in a bounded or periodic domain is often characterized by the evolution of its variance $\langle C^2 \rangle$, a quantity known to decrease with time when the flow is incompressible and no source of scalar is present \cite{Pierrehumbert1994}. The case of mixing in compressible flows has received less attention: Vergassola showed that a compressible flow
can modify scalar transport either in stationary cellular flows or random delta correlated flows \cite{Vergassola1997}; in the context of reactive flows, compressibility was proved to have a strong influence on chemical reactions or population growth \cite{perkelar2010,Pigolotti2012} because it controls the local density of fluid particles. Surprisingly, if most flows encountered in nature are incompressible, advection diffusion by an effective compressible flow field arises naturally when mixing large molecules, colloids, or living cells in moderate Reynolds number flows. 
Even in the case of flow tracers, the presence of background inhomogeneities will result in a drift velocity $\mathbf{V}_{\rm drift}$ between the fluid flow $\mathbf{V}(\mathbf{r},t)$ and the transported species. This situation is encountered in a large variety of situations: thermophoresis (Soret effect) leads to a drift velocity proportional to the temperature gradient \cite{soret1879,Braun2002}, diffusiophoresis is responsible for focusing and defocusing of colloids due to salt gradients \cite{anderson1989,Abecassis2009}, and chemotaxis allows living cells to move with a drift velocity proportional to the local gradient of food \cite{Munoz2010}. In these three cases, the drift term is generally not divergence free because the inhomogenous field does not satisfy the Laplace equation.\\ 
A recent experimental study showed how colloid mixing is tuned by diffusiophoresis in a chaotic flow \cite{Deseigne2013}. In this letter we demonstrate by means of numerical simulations that it is the compressible nature of the drift velocity which is responsible for this tuning, and that it may produce unexpected effects such
as increase of scalar variance at small times. 

We study the mixing of colloids and salt with respective concentrations $C(\mathbf{r},t)$ and $S(\mathbf{r},t)$ under the action of the velocity $\mathbf{V}(\mathbf{r},t)$. In the presence of salt gradients,  electrokinetic effects result in a drift velocity $\mathbf{V}_{\rm dp}=\alpha \nabla \log S$ between the colloids and the fluid motion, $\alpha$ being a constant with dimension of a diffusion coefficient whose magnitude and sign depend on the precise nature of the salt and colloids \cite{anderson1989,Abecassis2009}. Starting from a situation with initial concentration profiles $C_0(\mathbf{r})$ and $S_0(\mathbf{r})$, the time evolution of the concentrations is then given by the coupled advection diffusion equations: 
\begin{eqnarray}
&\frac{\displaystyle \partial S}{\displaystyle \partial t} + \nabla \cdot  S \mathbf{V} = D_s \nabla^2 S,\\
&\frac{\displaystyle \partial C}{\displaystyle \partial t} + \nabla \cdot C (\mathbf{V}+ \mathbf{V}_{{\rm dp}}) = D_c \nabla^2 C,\\
\label{eqadv2}
&\mathbf{V}_{{\rm dp}} = \alpha \nabla \log S,
%
\end{eqnarray}

 \begin{figure*}
\centering
\includegraphics[width=0.99\textwidth]{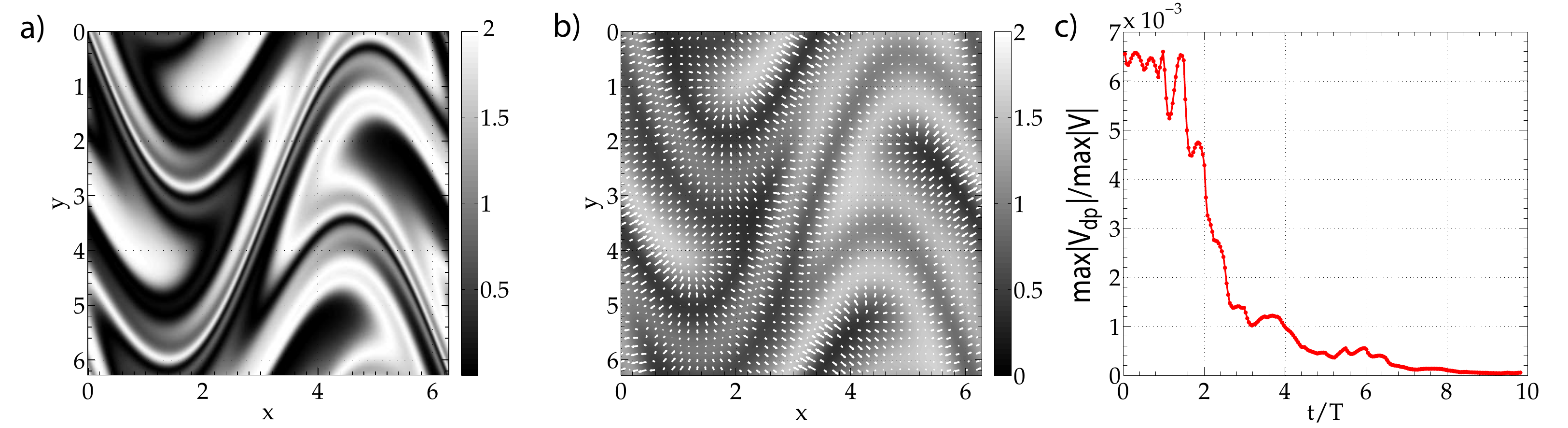}
\caption{a) Colloid concentration field at $t=5 T$ in the reference case ($D_c=10^{-4}$, $Pe=65000$) for an initial concentration $C_0=1+\sin(x)$. b) Salt concentration field at $t=5 T$ for $S_0=1.01+\sin(x)$ ($D_s=10^{-2}$, $Pe=650$) together with the corresponding drift velocity field $\mathbf{V}_{{\rm dp}} = \alpha \nabla \log S$ with $\alpha=0.1~D_s$. c) Corresponding time evolution of non dimensional maximum drift velocity, $\max |\mathbf{V}_{\rm dp} | / \max |\mathbf{V}|$.}
\label{setup} 
\end{figure*}

\noindent where $D_s$ and $D_c$ are respectively the diffusion coefficient of the salt and colloids. {In this situation, salt is mixed by the velocity field $\mathbf{V}$ independently of the colloids, while colloid concentration is also coupled to the salt concentration through the drift velocity $\mathbf{V}_{{\rm dp}}$.} In order to study the impact of the salt gradients onto colloid mixing, we restrict ourselves to 2D situations; we achieve chaotic mixing using a time periodic velocity field with chaotic streamlines, the so-called sine flow that has been a standard tool for studies of chaos in spatially smooth flows \cite{Pierrehumbert1994,muzzio1999,Ngan2011}. We recall it is $T$ periodic with two sub cycles for which the velocity field is $\mathbf{V}(\mathbf{r},t)=(\sin(y),0)$ for $nT\leq t < (n+1/2)T$ and  $\mathbf{V}(\mathbf{r},t)=(0,\sin(x))$ for $(n+1/2) T\leq t < (n+1)T$, which ensures the velocity field is divergence free at any time. For given initial concentration profiles  $C_0(\mathbf{r})$ and $S_0(\mathbf{r})$, equations (1) and (2) are solved with periodic boundary conditions using a pseudo spectral method with resolution $512^2$ for a square box of length $L=2 \pi$. We use an Adams-Bashford temporal scheme with order 2 and $dt=0.002$ to ensure scalar gradient and Laplacian are well resolved for diffusivities in the range $D_c \in [10^{-2},10^{-4}]$, which corresponds to Peclet numbers in the range $\text{Pe}= \max |\mathbf{V}| L/D_c\in [0.065, 6.5] 10^4$.  For all simulations we set $T=0.8~L/\max |\mathbf{V}|$; in this regime of non global chaos {(see Poincar\'e sections in figures 3-10 of reference \cite{handbookmix})}, as in many natural situations, one can distinguish between thin structures developing due to stretching and folding, and poor mixing in regular regions which govern the long time decay of scalar variance \cite{Pierrehumbert1994}. This is visible in figure \ref{setup}(a) which displays the colloid concentration field obtained in the reference case $\mathbf{V}_{\rm dp}=\mathbf{0}$ for an initial profile $C_0(\mathbf{r})=1+\sin(x)$ after 5 cycles of mixing.\\
{When diffusiophoresis comes into play, the drift term $\mathbf{V}_{\rm dp}=\alpha \nabla \log S$ depends on the instantaneous salt gradients. {In order to be consistent with experiments \cite{Deseigne2013,Abecassis2009}, we set $\alpha=0.1\,D_s$ and $D_s=0.01$ for all simulations. We define the initial salt concentration profile $S_0=1+\sin(x)+s$, $s=0.01$ being an offset accounting for the unavoidable homogeneous background concentration of ionic species (buffer solution).} As observed in experiments \cite{Deseigne2013}, one expects the drift term to modify colloid mixing depending on whether the initial colloid concentration field $C_0(x)$ is correlated, or anti-correlated, with the initial salt concentration profile $S_0(x)$. {For all Peclet numbers} we study the evolution of colloid concentration variance $\langle (C-\langle C \rangle)^2 \rangle$ for two different initial profiles $C_0=1\pm \sin(x)$, and compare the results to the reference case $\alpha=0$. 
The drift velocity, which does not depend on $C(\mathbf{r},t)$, is displayed in figure \ref{setup}(b) together with the corresponding salt concentration field  computed at $t/T=5$. With the chosen parameters, the modification of the total velocity field acting on the colloids, $\mathbf{V}+\mathbf{V}_{\rm dp}$, is never larger than $1\%$ as shown in figure \ref{setup}(c). However, we will demonstrate that this small variation leads to a strong alteration of mixing because $\mathbf{V}_{\rm dp}$ is not divergence free (figure \ref{setup}(b)).}

\begin{figure}[h]
  \begin{center}
  \includegraphics[width=0.9\columnwidth]{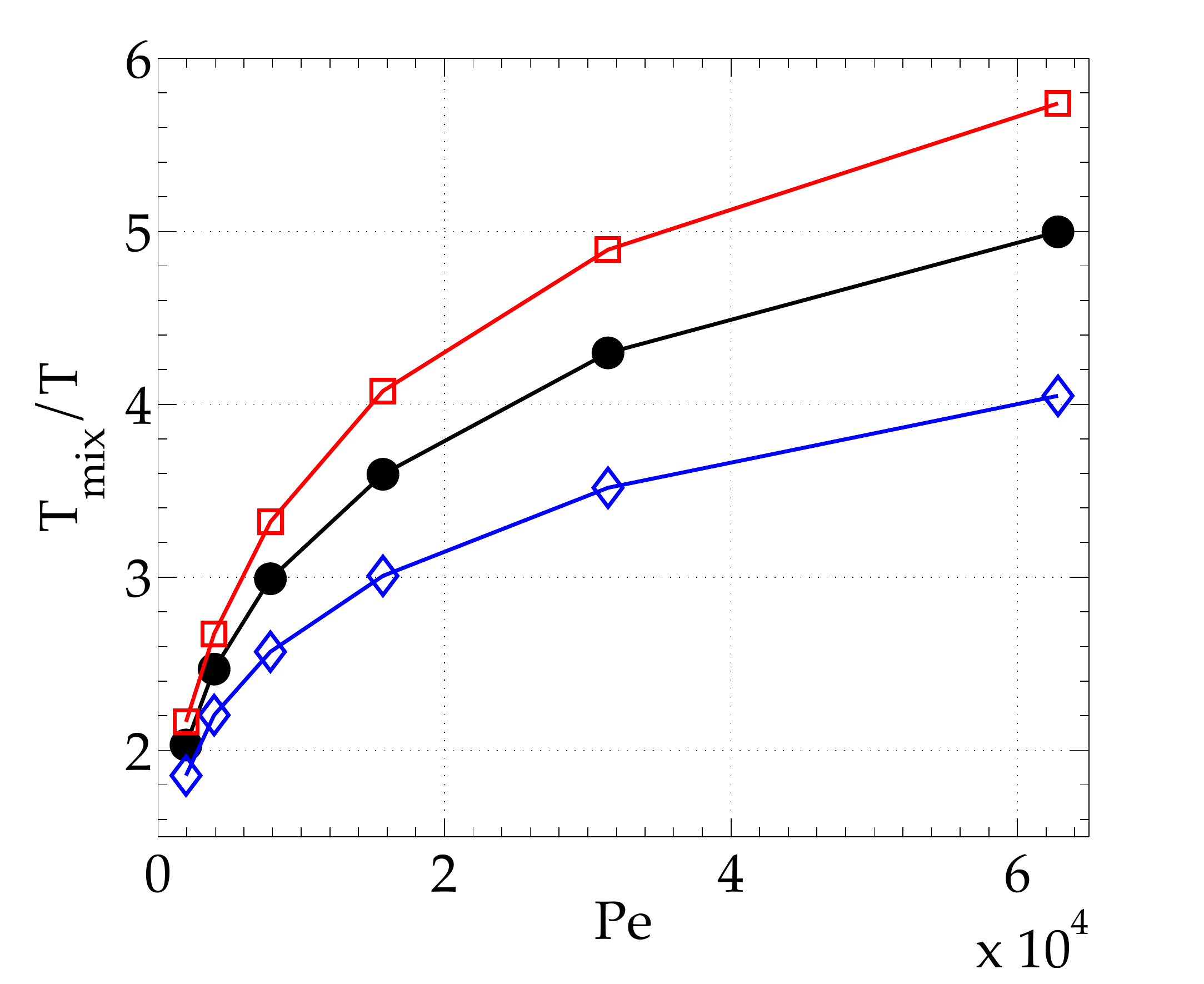}
    \caption{(colour online) Mixing time $T_\text{mix}/T$ as a function of the Peclet number $\text{Pe}= \max |\mathbf{V}| L/D_c$ as measured from time evolution of the variance $\sigma^2=\langle C^2 \rangle -\langle C \rangle^2$ with and without diffusiophoresis. For all cases $S_0=1.01+\sin x$ with $D_s=10^{-2}$ and $D_c=1-100D_s$. {\Large $\bullet$}: reference case with no diffusiophoresis  $C_0=1+\sin(x)$ and $\alpha=0$. $\square$: hypo diffusive case $C_0=1+\sin x$, $\alpha=0.1~D_s$. {\Large $\diamond$}: hyper diffusive case $C_0=1-\sin(x)$, $\alpha=0.1~D_s$.}
  \label{tmixpeclet}
  \end{center}
\end{figure}

{In figure \ref{tmixpeclet}, we display the evolution of the mixing time, $T_\text{mix}$, needed to decrease the initial colloid concentration variance by a factor $2$, as a function of the Peclet number $\text{Pe}= \max |\mathbf{V}| L/D_c$.  As already observed in experiments \cite{Deseigne2013}, when salt and colloid are initially injected together (Salt-in configuration, $C_0(\mathbf{r}) = 1 + \sin x \sim S_0(\mathbf{r})$, thereafter called hypo diffusive), mixing is approximately $15\%$ slower than in the reference case $\alpha=0$.  On the contrary when salt and colloids are injected separately (Salt-out configuration, $C_0(\mathbf{r}) =1 - \sin x$ {and $S_0(\mathbf{r}) =1.01 + \sin x$}, thereafter called hyper diffusive), the mixing is nearly $20\%$ faster. 
This result supports the idea that the action of diffusiophoresis can be seen as a modification of colloid transport properties through an effective diffusivity $D_\text{eff}$ corresponding to an effective Peclet number $\text{Pe}= \max |\mathbf{V}| L/D_\text{eff}$. Using the curves $T_\text{mix}(\text{Pe})$, well fitted by a power law $T_\text{mix}/T \propto \text{Pe}^{1/4}$ for the reference case, one would then find $\text{Pe}_{\text{eff}} \sim 1.5 ~\text{Pe}$ and $\text{Pe}_{\text{eff}} \sim 0.5 ~\text{Pe}$ for the hypo diffusive (salt-in) and hyper diffusive (salt-out) configurations respectively. This means that a correction to the velocity field smaller than $1\%$ leads to a $50\%$ change in the colloid effective diffusivity!}

\begin{figure}[h]
  \begin{center}
  \includegraphics[width=0.9\columnwidth]{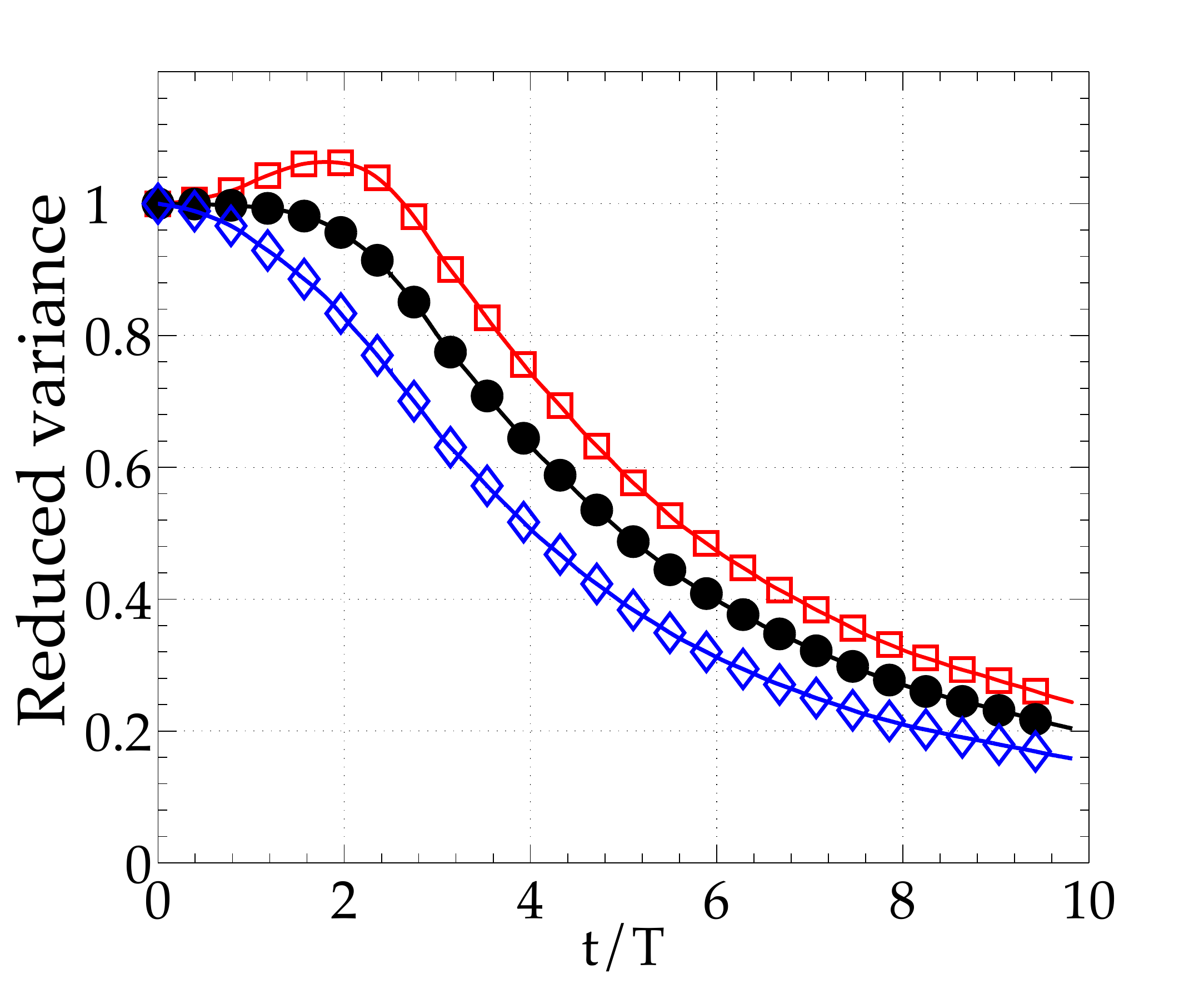}
    \caption{(colour online) Time evolution of non dimensional colloid concentration variance {$\sigma^2(t)/\sigma^2(0)$ with $\sigma^2(t)=\langle C^2 \rangle-\langle C \rangle^2$} and $S_0=1.01+\sin(x)$, $D_s=10^{-2}$, $D_c=10^{-4}$. {\Large $\bullet$}:reference case with no diffusiophoresis  $C_0=1+\sin x$, $\alpha=0$. $\square$: hypo diffusive case $C_0=1+\sin x$, $\alpha=0.001$. {\Large $\diamond$}: hyper diffusive case $C_0=1-\sin x$, $\alpha=0.001$.} 
  \label{variancetime}
  \end{center}
\end{figure}

{However, this mean field approach, if interesting for interpreting the long time behavior, may not catch the short-time evolution of colloid variance
when {concentration} gradients are not generated yet, so that diffusion is negligible. As displayed in figure \ref{variancetime}, if one always observes a decrease of scalar variance at small times in the hyper diffusive and reference cases, the hypo diffusive case is more appealing: the variance first increases for $t \leq 2$ before decreasing for $t \geq 2$. This short-time increase is incompatible with an effective diffusivity which would predict a variance decreasing with rate $\text{d} \langle C^2 \rangle/\text{d}t = - 2 D_{\rm eff} \langle(\nabla C)^2\rangle$, where $\langle \cdot \rangle$ stands for spatial averaging over one period of the flow. For a better understanding of the colloid variance evolution, one may then come back to equation (2), and derive an equation for the scalar energy $C^2$ valid for compressible flows. One obtains:}

\begin{eqnarray*}
& \frac{\displaystyle 1}{\displaystyle 2}\frac{\displaystyle \partial C^2 }{\displaystyle \partial t} + \nabla \cdot \left( (\mathbf{V}+\mathbf{V}_{{\rm dp}}) \frac{\displaystyle C^2}{\displaystyle 2} - D_c C \nabla C \right)= \\
&- D_c (\nabla C)^2 - \frac{\displaystyle C^2}{\displaystyle 2} \nabla \cdot \mathbf{V}_{{\rm dp}}
\end{eqnarray*}

{This equation, although similar to the classical scalar energy budget, contains an additional term proportional to the drift velocity field compressibility $\nabla \cdot \mathbf{V}_{\rm dp}$ that does not vanish in the present case. When averaging over one flow period, all terms written as a divergence disappear and one obtains a new global scalar energy budget valid for compressible flows:
\begin{eqnarray}
&\frac{\displaystyle  1}{\displaystyle  2}\frac{\displaystyle \text{d}\langle C^2 \rangle}{\displaystyle \text{d} t} = - D_c \langle (\nabla C)^2 \rangle - \langle \frac{\displaystyle C^2}{\displaystyle 2} \nabla \cdot \mathbf{V}_{\rm {dp}} \rangle.
\end{eqnarray}
}

\begin{figure}[h]
  \begin{center}
  \includegraphics[width=0.9\columnwidth]{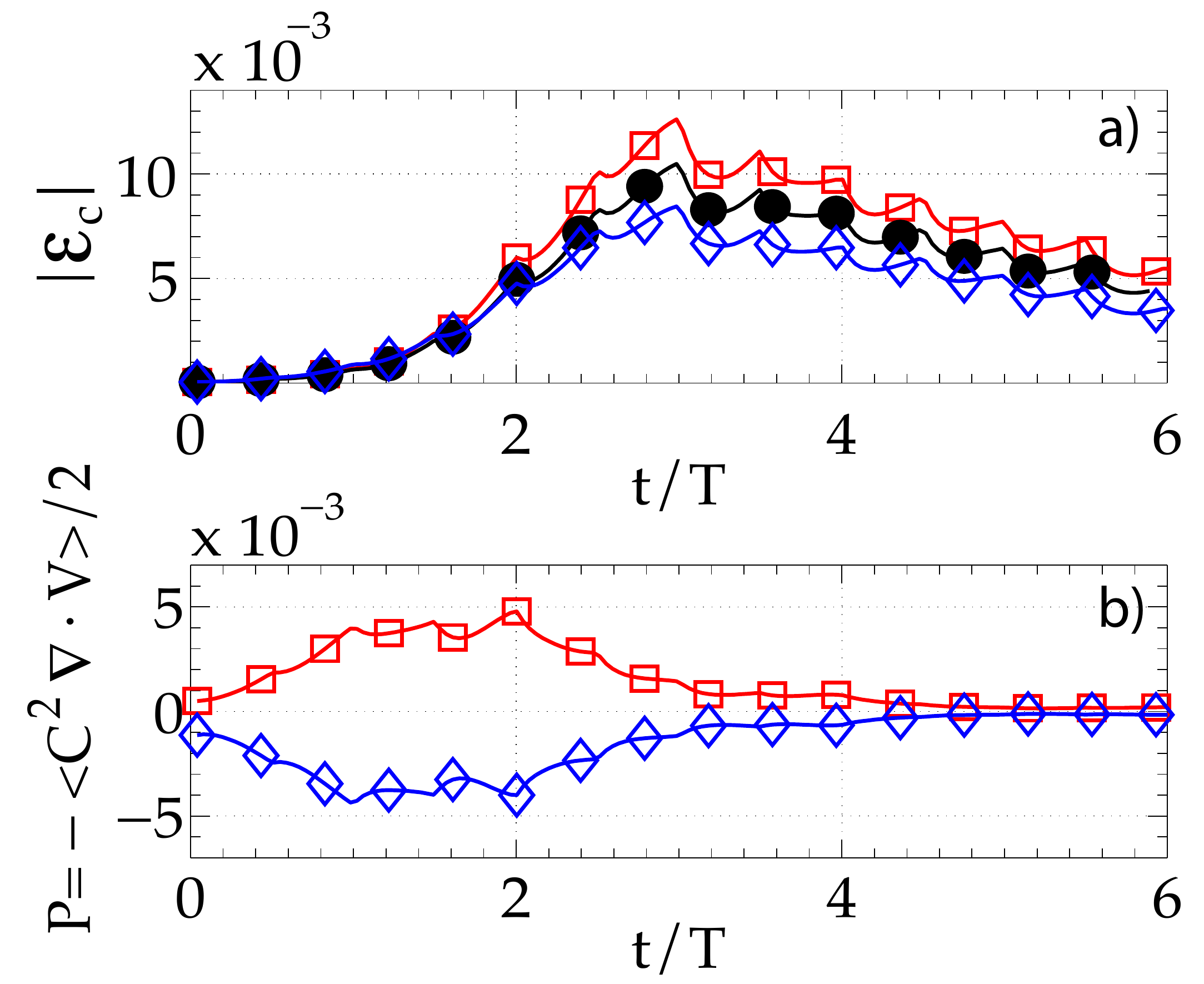}
    \caption{(colour online) (a) Time evolution of colloid concentration dissipation $|\varepsilon_c|=D_c \langle (\nabla C)^2 \rangle$. {\Large $\bullet$}: reference case with no diffusiophoresis. $\square$: hypo diffusive (salt-in) case. {\Large $\diamond$}: hyper diffusive (salt out) case. (b) Corresponding time evolution of production term $P = - \langle C^2 \nabla \cdot \mathbf{V}_{\rm dp} \rangle/2$.}
  \label{scaldiss}
  \end{center}
\end{figure}

{Because mean scalar concentration remains conserved even in the case of compressible flows, this equation also {gives the evolution of scalar variance} $\sigma^2=\langle C^2 \rangle-\langle C \rangle^2$. Therefore scalar variance in compressible flows results in the competition between scalar dissipation $\varepsilon_c=- D_c \langle (\nabla C)^2 \rangle$ and production $P=- \langle C^2\nabla \cdot \mathbf{V}_{\rm dp} \rangle/2$, the later being proportional to the mutual coherence between $C^2$ and the total flow field compressibility $\nabla \cdot \mathbf{V}_{\rm dp}$.} This equation, well verified in our case at any time step, helps understanding the evolution of scalar variance in the presence of diffusiophoresis. Indeed, as seen in figure \ref{scaldiss}(a), dissipation alone can not explain the observed differences in scalar variance at short times because $\varepsilon_c$ remains very weak until small scale fluctuations have been created by stretching and folding. For the two first mixing cycles $t/T \leq 2$, the evolution of the variance is then governed by the production term $\displaystyle P= - \alpha \langle \displaystyle C^2 \nabla^2 \log S \rangle/2$, positive for the salt-in configuration and negative for the salt-out configuration as demonstrated in the figure \ref{scaldiss}(b). {This effect disappears when the salt has been mixed. Considering the Peclet number of the salt ($Pe=650$), figure \ref{tmixpeclet} shows the mixing time is of the order of two mixing cycles, which means salt is totally mixed at $t=2 T_{\rm mix} \sim 4 T$.} 
For $t \geq 4T$ both the drift velocity field and the production term have then become very weak (figure \ref{setup}(c) and \ref{scaldiss}), so that the decay of variance is mainly governed by scalar dissipation thereafter. We may then conclude that diffusiophoresis is globally impacting mixing because compressibility modifies the colloid concentration field at small times while regular mixing of this modified initial condition follows shifted in time.

\begin{figure}[h]
  \begin{center}
  \includegraphics[width=1\columnwidth]{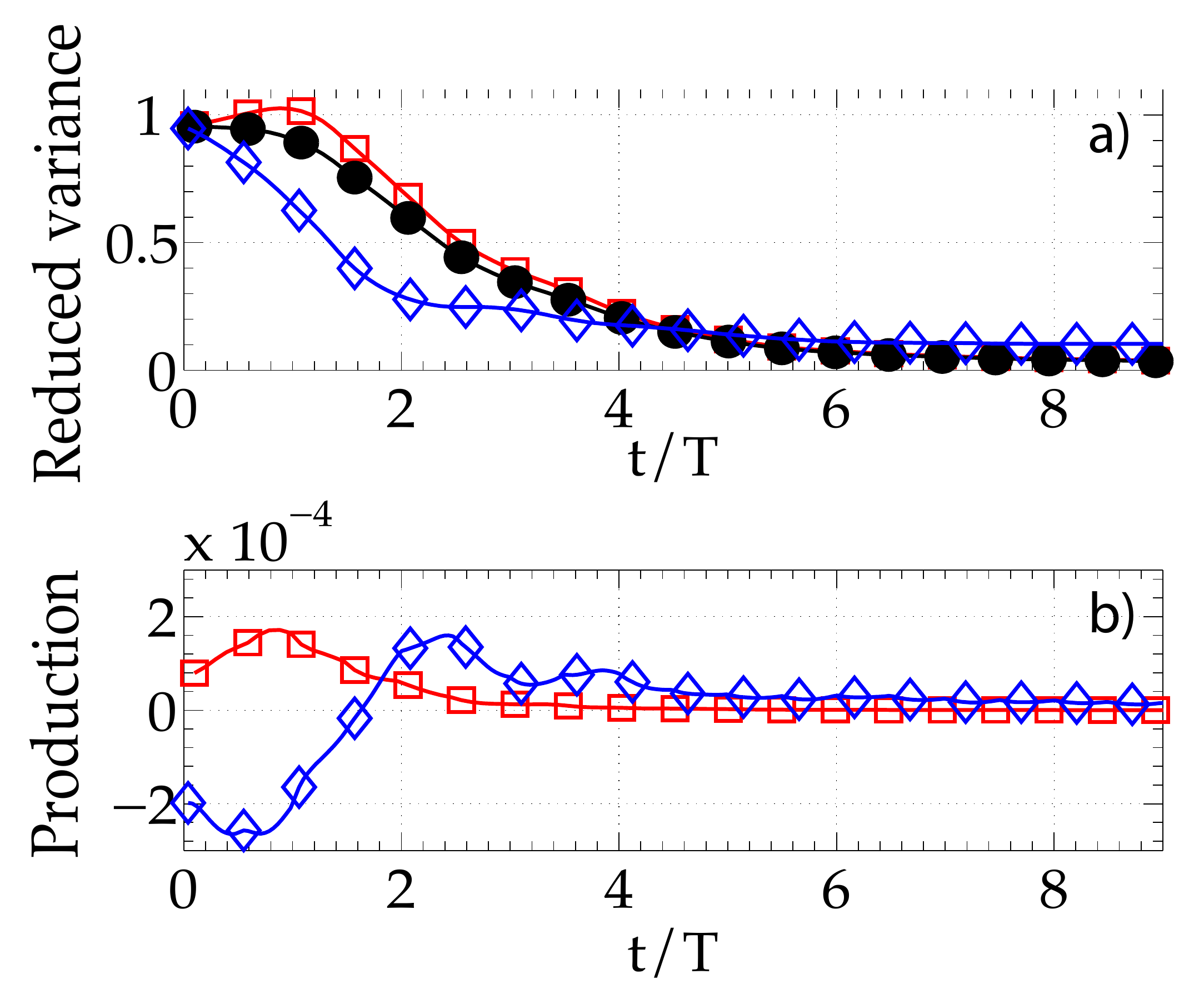}
    \caption{(colour online) (a) Time evolution of non dimensional colloid concentration variance for a patch of salt $S_0=0.01+\exp\left(-4(x^2+y^2)\right)$ with $D_s=10^{-2}$, $D_c=10^{-4}$. {\Large $\bullet$}: reference case with no diffusiophoresis  $C_0=\exp\left(-4(x^2+y^2)\right)$, $\alpha=0$. $\square$: hypo diffusive case $C_0=\exp\left(-4(x^2+y^2)\right)$. {\Large $\diamond$}:hyper diffusive case $C_0=1-\exp\left(-4(x^2+y^2)\right)$. For the last two cases $\alpha=0.001$. (b) Corresponding time evolution of production term $P = - \langle C^2 \nabla \cdot \mathbf{V}_{\rm dp} \rangle/2$.}
  \label{variancegauss}
  \end{center}
\end{figure}

{Up to now we have dealt with modal $(\sin x)$ and therefore $``$non localized$"$ initial conditions for $C_0$ and $S_0$. However since mixing in the presence of diffusiophoresis is governed by the coherence between salt and colloids concentration, it strongly depends on initial conditions. Therefore it is interesting to look at more realistic situations where the initial condition is more $``$localized$"$, such as in a patch of fluid. For instance, one can compare the cases when an initial patch of salt is introduced in a micro mixer already filled with well mixed colloids, or else when a patch of salt+colloids is injected in pure water.} This situation can be reproduced with gaussian profiles $S_0=\exp\left(-4(x^2+y^2)\right)$ for which the salt-in and salt-out configurations would correspond to $C_0=\exp\left(-4(x^2+y^2)\right)$ and $C_0=1-\exp\left(-4(x^2+y^2)\right)$ respectively. For these initial concentration profiles, if the reference and salt-in configurations lead to classical evolutions for the variance as shown in figure \ref{variancegauss}(a), one observes an unexpected evolution with two time scales for the salt-out configuration. After a very fast mixing at short times $t \leq 2$, colloids  start to mix at reduced speed so that finally the salt-out configuration is less efficient for mixing on long time scales. This again can be understood by looking at the production term. If $P$ remains always positive for the hypo diffusive case as shown in figure \ref{variancegauss}(b), it is negative at short times for the hyper diffusive case before becoming positive for $t \geq 1.8$. This is because in the case of the {hyper diffusive} configuration, the large amount of colloids with homogenous concentration around the patch of salt is being unmixed. This unmixing reinforces the variance production at long times, resulting in a much slower mixing than without diffusiophoresis.

{To conclude, we have studied 2D chaotic mixing of colloids under the action of diffusiophoresis which produces a velocity drift and makes colloids no longer be tracers of the flow motion. We have demonstrated that this very small drift, that leads to a correction to the velocity field smaller than $1\%$, is responsible for an important change in mixing because it strongly modifies the topology of the total flow, which is no longer divergence free. This compressibility is at the origin of the scalar variance production term $P=- \langle C^2 \nabla \cdot \mathbf{V}_{\rm dp} \rangle/2$, necessary to understand: i) the evolution of scalar variance at small times, not explained in the framework of effective diffusion for the hypo diffusive case. ii) acceleration or delay of mixing depending on the coherence between salt and colloids concentration fields. One remarkable property of this compressible effects is the capacity to unmix an initially homogeneous colloidal solution by simply adding salt gradients. If the present results were obtained on the physical case of colloid mixing in the presence of chemical gradients, they are more general and may play an important role in cell dynamics because the total flow is compressible \cite{Munoz2010}. Effective flow compressibility should also apply to temperature gradients (Soret effect), and may help to understand older experiments of DNA trapping and amplification in laminar thermally driven flows  \cite{Braun2002}, the Laplacian of the temperature field being an image of the local DNA concentration. Finally, we note these compressibility effects are not a manifestation of laminar mixing only. {They are also observed in particle laden turbulent flows providing the particles do not follow the fluid motions because they have inertia \cite{bib:maxey1987_JFM,bib:shaw2003}, or because they are sensitive to gravitational field \cite{Durham2013}. Bridging between the different results, this suggests the possibility of a common frame of description for turbulent clustering and diffusiophoretic mechanisms via compressible effects.}} 

This collaborative work was supported by the LABEX iMUST (ANR-10-LABX-0064) of Universit\'e de Lyon,
within the program $``$Investissements d'Avenir$"$ (ANR-11-IDEX-0007) operated by the French
National Research Agency (ANR).

\bibliography{main}

\end{document}